\documentclass[print,authoryear,12pt]{elsarticle}
\usepackage{amsmath}
\usepackage{amssymb}
\usepackage{amstext}
\usepackage{amsthm}
\usepackage{bm}
\usepackage{color}
\usepackage{epstopdf}
\usepackage{graphicx}
\usepackage{mathrsfs}
\usepackage{multirow}
\usepackage{natbib}
\usepackage{pdfsync}
\usepackage{rotating}

\journal{Computational Statistics \& Data Analysis}

\newcommand{\ve}[1]{\bm{{#1}}}
\newcommand{\vesub}[2]{\bm{{#1}}_{#2}}
\newcommand{\vesup}[2]{\bm{{#1}}^{#2}}
\newcommand{\vess}[3]{\bm{{#1}}_{#2}^{#3}}
\newcommand{\hve}[1]{\hat{\ve{#1}}}
\newcommand{\hvesub}[2]{\hat{\ve{#1}}_{#2}}

\newcommand{\hvess}[3]{\hat{\ve{#1}}_{#2}^{#3}}

\begin{document}
\begin{frontmatter}
\title{Robust functional ANOVA model with t-process}
\author[a]{Chen Zhang}
\address[a]{Department of Statistics and
Finance, University of Science and Technology of China, Hefei, China.}
\author[a]{Zimu Chen}
\author[a]{Zhanfeng Wang
\footnote[1]{Corresponding author: Department of Statistics and
Finance, University of Science and Technology of China, Hefei
230026, China. E-mail: zfw@ustc.edu.cn}}
\author[a]{Yaohua Wu}

\begin{abstract}
Robust estimation approaches are of fundamental importance for statistical modelling. To reduce susceptibility to outliers, we propose a robust estimation procedure with t-process  under functional ANOVA model. Besides common mean structure of the studied subjects, their personal characters are also informative, especially for prediction. We develop a prediction method to predict the individual effect. Statistical properties, such as robustness and information consistency, are studied. Numerical studies including simulation and real data examples show that the proposed method performs well.
\end{abstract}


\end{frontmatter}

\setcounter{section}{0}
\setcounter{equation}{0}
\numberwithin{equation}{section}

\section{Introduction}

The analysis of variance (ANOVA) functional models are frequently applied to analyze data in many fields, such as econometrics, sociology and medical follow-up studies, such as application to the breast cancer data \citep{r7}, the angles data of the hip over walking cycles \citep{r12}, the vehicle road load data acquisition computer model \citep{r3} and so on.

Functional ANOVA model is an adaptation of linear model or analysis of variance to analyze functional observations, which are also widely studied, such as \cite{r19}, \cite{r6}, \cite{r2} and therein reference. To the best of our knowledge, functional ANOVA models in literature rarely consider individual effect of the studied subject (each observed functional curve). However, for market penetration of new products \citep{r1}, data are collected from 86 countries in 7 regions. Market penetrations from different countries, such as the developing and developed countries, have dissimilar growth trend, because of discrepant economical levels. Hence, market penetration of each country has personal character besides the common mean structure. This paper not only considers the main structure of market penetration, but also focuses on individual effect of each country via a nonparametric random effect. Without loss of generality, we consider a one-way functional ANOVA concurrent regression model,
\begin{equation}\label{1.1}
y_{ij}(t)=\mu(t)+\alpha_i(t)+\tau_{ij}(\vesub{u}{ij}(t))+\varepsilon_{ij}(t), ~j=1,\ldots,J_i,~ i=1,\ldots,I,
\end{equation}
where $t$ is the observed time, $y_{ij}(t)$ is a functional response for batch $j$ under level $i$, $\mu(t)$ is a grand function, $\alpha_i(t)$ measures the main effect of level $i$, $\tau_{ij}$ is a nonparametric random effect depicting personal character for the $j$-th subject at level $i$, $\vesub{u}{ij}(t)$ is a functional covariate with dimension $p$, and $\varepsilon_{ij}(t)$ is an error term. For identification of $\alpha_i(\cdot)$, it needs
\[\sum\limits_{i=1}^I{\alpha_i(\cdot)}=0.\]

When random effect $\tau_{ij}=0$ and $\varepsilon_{ij}(t)$ has Gaussian distribution, model (\ref{1.1}) becomes the common one way functional ANOVA model. For example, \cite{r2} and \cite{r7}, respectively, used smoothing spline model and  B-splines method to build this functional ANOVA model; \cite{r12} studied this model based on unequally sampled noisy curves.

When $\tau_{ij}$ follows Gaussian process (GP) with mean function 0 and kernel function $k_{ij}$, denoted by $GP(0, k_{ij})$ and $\varepsilon_{ij}(t)$ has Gaussian distribution, we denote model (\ref{1.1}) by Gaussian process ANOVA model, saying GP model. Section 5.5 in \cite{r16} mentions the GP model. However, the GP is sensitive to outliers. For robustness, this paper assumes  $\varepsilon_{ij}$ has heavy-tail process, such as t-process. For theoretical studies and application of heavy-tail processes, please see \cite{r21}, \cite{r22}, \cite{r15} and \cite{r18} and so on. To illustrate the proposed method, this paper employs an extended t-process (ETP)  \citep{r18} to depict the error term. Our method can be directly extended to other t-process.

This paper combines t-process error $\varepsilon_{ij}$ with Gaussian process random effect $\tau_{ij}$ in model (\ref{1.1}), which leads to a robust functional ANOVA model. Independence of random effect and error term in this paper makes computation procedure more complicated compared to \cite{r18} which uses a joint ETP process to build their model. We propose a computation method based on Gaussian approximation. Prediction of random effect is also an important for studying personal character. We develop a prediction method for functional ANOVA model. Statistical properties of the estimation method, such as robustness and information consistence, are investigated. Numerical studies including simulation and real examples show that the proposed method has robustness against outliers.

The remainder of this paper is organized as follows. Section 2 proposes a robust functional ANOVA model with t-process, and builds prediction and parameter estimation procedure. Asymptotic properties of estimation method are discussed in Section 3. Simulation studies and real examples are reported in Section 4, followed by the conclusion in Section 5. All the proofs are in Appendix.

\setcounter{equation}{0}

\section{Robust functional ANOVA model}

Considering model (\ref{1.1}), we assume that $\tau_{ij}(\vesub{u}{ij}(\cdot))\sim GP(0,k_{ij})$, where $k_{ij}$ is a kernel function. Hereafter, for convenience of notations, let $\tau_{ij}(\cdot)=\tau_{ij}(\vesub{u}{ij}(\cdot))$. For robustness to outlier, let $\varepsilon_{ij}$ be from the extended t-process $ETP(\nu,\nu-1,0,\sigma^2 k_\varepsilon)$ with $k_\varepsilon(u,v)=I(u=v)$ for $u,v \in R$ \citep{r18}, where evaluated values of $\ve{z}=(\varepsilon_{ij}(t_1),...,\varepsilon_{ij}(t_n))^\top$ at any data points $t_1,\ldots,t_n$, follows an extended multivariate t-distribution (EMTD) with the density function
\[
f(\ve{z})=|2\pi(\nu-1)\sigma^2\vesub{I}{n}|^{-\frac{1}{2}}
\frac{\Gamma(n/2+\nu)}{\Gamma(\nu)}
\left(1+\frac{||\ve{z}-\ve{h}||^2}
{2(\nu-1)\sigma^2}\right)^{-(n/2+\nu)}.
\]

Denoted by $\ve{\beta}(t)=(\mu(t),\alpha_1(t),\ldots,\alpha_I(t))^\top$, $\ve{z}_{ij}=(1,0,\ldots,0,1,0,\ldots,0)^\top$ where the first and $(i+1)$-th elements are 1 and the others are 0. Model (\ref{1.1}) is rewritten as
\begin{equation}\label{2.1}
y_{ij}(t)=\vess{z}{ij}{\top}\ve{\beta}(t)+\tau_{ij}(t)+\varepsilon_{ij}(t).
\end{equation}
From this model, it follows that
\begin{equation}\label{2.2}
\begin{aligned}
y_{ij}|\tau_{ij},\sigma^2,\nu &\sim ETP(\nu,\nu-1,\vess{z}{ij}{\top}\ve{\beta}+\tau_{ij},\sigma^2k_\varepsilon),\\
\tau_{ij} &\sim GP(0,k_{ij}).
\end{aligned}
\end{equation}

\subsection{Prediction}

Let $\vesub{Y}{ij}=(y_{ij}(t_1),\ldots,y_{ij}(t_n))^\top$ be observed data for the $j-$th curve under level $i$, and $\vesub{\tau}{ij}=(\tau_{ij}(t_1),\ldots,\tau_{ij}(t_n))^\top$ at the observed times $\ve{t}=\{t_1,\ldots,t_n\}$. Denoted by $\ve{\mathcal{D}_n}=\{\vesub{Y}{ij},j=1,\ldots,J_i,i=1,...,I\}$. For a new point $t^*$, $E(y_{ij}(t^*)|\ve{\mathcal{D}_n})$ is used to predict $y_{ij}$ at point $t^*$, denoted by $\hat{y}_{ij}(t^*)$. From model (\ref{2.2}), we show that
\begin{equation}\label{2.3}
\begin{aligned}
E(y_{ij}(t^*)|\ve{\mathcal{D}_n})&=E(E(y_{ij}(t^*)|\tau_{ij}(t^*),\ve{\mathcal{D}_n}))\\
&=\int{\{\vess{z}{ij}{\top}\ve{\beta}(t)+\tau_{ij}(t^*)\}p(\tau_{ij}(t^*)|
\ve{\mathcal{D}_n})d\tau_{ij}(t^*)}.
\end{aligned}
\end{equation}

Since the conditional distribution of $\tau_{ij}(t^*)$ for given $\ve{\mathcal{D}_n}$ is involved in intractable integration, (\ref{2.3}) is complicated to be computed. Gaussian approximation  method is employed to calculate (\ref{2.3}). It follows that
\begin{equation}\label{2.4}
\begin{aligned}
p(\tau_{ij}(t^*)|\ve{\mathcal{D}_n})&=\int{p(\tau_{ij}(t^*)|\vesub{\tau}{ij},\ve{\mathcal{D}_n})
p(\vesub{\tau}{ij}|\ve{\mathcal{D}_n})d\vesub{\tau}{ij}}\\
&\approx\int{p(\tau_{ij}(t^*)|\vesub{\tau}{ij})
p_G(\vesub{\tau}{ij}|\ve{\mathcal{D}_n})d\vesub{\tau}{ij}},
\end{aligned}
\end{equation}
where $p_G(\vesub{\tau}{ij}|\ve{\mathcal{D}_n})$ is the Gaussian approximation of $p(\vesub{\tau}{ij}|\ve{\mathcal{D}_n})$ (details seen in Section 2.2).

Since $\tau_{ij} \sim GP(0,k_{ij})$, we have
\[\begin{pmatrix}
\vesub{\tau}{ij}\\
\tau_{ij}(t^*)
\end{pmatrix}
\sim N
\begin{pmatrix}
0,
\begin{pmatrix}
\vesub{K}{ij}&\vesub{k}{ij}(t^*,\ve{t})\\
\vesub{k}{ij}(t^*,\ve{t})^\top&k_{ij}(t^*,t^*)
\end{pmatrix}
\end{pmatrix},\]
where $\vesub{K}{ij}=(k_{ij}(t_k,t_l))_{n\times n}$ is the covariance matrix of $\vesub{\tau}{ij}$, $\vesub{k}{ij}(t^*,\ve{t})=(k_{ij}(t^*,t_1),...,k_{ij}(t^*,t_n))^\top$. It follows that $p(\tau_{ij}(t^*)|\vesub{\tau}{ij})$ is normal distribution with mean
$\vesup{a}{\top}\vesub{\tau}{ij}=\vesub{k}{ij}(t^*,\ve{t})^\top\vess{K}{ij}{-1}\vesub{\tau}{ij}$ and variance
$\sigma^{*2}=k_{ij}(t^*,t^*)-\vesub{k}{ij}(t^*,\ve{t})^\top\vess{K} {ij}{-1}\vesub{k}{ij}(t^*,\ve{t})$.
From Section 2.2, we have
\[p_G(\vesub{\tau}{ij}|\ve{\mathcal{D}_n})=N(\hvesub{\tau}{ij},~\ve{\Omega}),~~
\ve{\Omega}\triangleq(\vess{K}{ij}{-1}+\vesub{D}{m})^{-1},\]
where $\hvesub{\tau}{ij}$ and $\vesub{D}{m}$ are defined in the next subsection.

It easily shows that $p(\tau_{ij}(t^*)|\ve{\mathcal{D}_n})$ has a normal density function $N(\vesup{a}{\top}\hvesub{\tau}{ij},\\\vesup{a}{\top}\ve{\Omega} \ve{a}+\sigma^{*2}),$ which indicates that ${y}_{ij}(t^*)$ can be estimated by
$$\hat{y}_{ij}(t^*)=E(y_{ij}(t^*)|\ve{\mathcal{D}_n})=\vess{z}{ij}{\top}\ve{\beta}(t)+\vesup{a}{\top}\hvesub{\tau}{ij}.$$
We know that
\begin{align}\label{2.5}
&\text{var}(y_{ij}(t^*)|\ve{\mathcal{D}_n})\nonumber\\
=&E(\text{var}(y_{ij}(t^*)|\tau_{ij}(t^*))|\ve{\mathcal{D}_n}))+
\text{var}(E(y_{ij}(t^*)|\tau_{ij}(t^*))|\ve{\mathcal{D}_n})).
\end{align}
Easily, we show
\begin{align}
&\text{var}(E(y_{ij}(t^*)|\tau_{ij}(t^*))|\ve{\mathcal{D}_n}))=\vesup{a}{\top}\ve{\Omega}\ve{a}+\sigma^{*2},\nonumber\\
&E(\text{var}(y_{ij}(t^*)|\tau_{ij}(t^*))|\ve{\mathcal{D}_n}))=\sigma^2.\nonumber
\end{align}
Hence, $\text{var}(y_{ij}(t^*)|\ve{\mathcal{D}_n})=\vesup{a}{\top}\ve{\Omega}\ve{a}+\sigma^{*2}+\sigma^2$ is used to estimate the variance of $\hat{y}_{ij}(t^*)$.

\subsection{Parameter estimation}

To apply the proposed method, it is necessary to estimate the unknown kernel function $k_{ij}(\cdot,\cdot)$, $\sigma^2$ and the function parameter $\ve\beta$.
For kernel function, a function family such as a squared exponential kernel and Mat\'{e}rn class kernel can be applied \citep{r16}. This paper takes a combination of a square exponential kernel and a non-stationary linear kernel,
\begin{align}
&k_{ij}(t_1,t_2)=k(t_1,t_2;\vesub{\theta}{ij})\nonumber\\
=&\theta_{ij0}\text{exp}\left\{-
\frac{\sum_{k=1}^p\theta_{ijk}(u_{ijk}(t_1)-u_{ijk}(t_2))^2}{2}
\right\}+\sum_{k=1}^p\eta_{ijk}u_{ijk}(t_1)u_{ijk}(t_2),\nonumber
\end{align}
where $\vesub{\theta}{ij}=(\theta_{ij0},\theta_{ij1},...,\theta_{ijp},\eta_{ij1},...,\eta_{ijp})^\top$ is hyper-parameter. We use B-spline to approximate the functional parameter $\ve{\beta}(t)$, that is $\ve{\beta}(t)=\ve{B}^\top\ve{\Phi}(t)$, where $\ve{\Phi}(t)=(\phi_1(t),\ldots,\phi_L(t))^\top, \{\phi_k(t),k=1,\ldots,L\}$ is a set of known basis functions and $\ve{B}$ is an unknown coefficient matrix with dimension $L\times(I+1)$.

Let $\ve{\theta}=(\vess{\theta}{11}{\top},\ldots,\vess{\theta}{IJ_I}{\top},\sigma^2)^\top,$
$\vesub{\Phi}{n}=(\ve{\Phi}(t_1),\ldots,\ve{\Phi}(t_n))_{L\times n}$,
$\vesub{X}{ij}=(1,0,\ldots,$ $0,1,0,\ldots,0)^\top$ with the first and $(i+1)$-th elements are 1 and the others are 0.
Then, we have
\begin{align}
\vesub{Y}{ij}|\vesub{\tau}{ij},\ve{B},\sigma^2,\nu &\sim EMTD(\nu,\nu-1,(\vess{X}{ij}{\top}\otimes\vess{\Phi}{n}{\top}) \text{Vec}(\ve{B})+\vesub{\tau}{ij},\sigma^2\ve{I_n}),\nonumber\\
\vesub{\tau}{ij}|\vesub{\theta}{ij} &\sim MVN(0,\vesub{K}{ij}),\nonumber
\end{align}
where $\otimes$ represents the Kronecker product. Then a marginal log-likelihood is
\begin{align}
\text{log~}p(\ve{\mathcal{D}_n}|\ve{\theta},\ve{B},\nu)&=\sum\limits_{i,j}{\text{log}\{p(\vesub{Y}{ij}|\vesub{\theta}{ij},
\ve{B},\sigma^2,\nu)\}}\nonumber\\
&=\sum\limits_{i,j}{\text{log}\int{p(\vesub{Y}{ij}|\vesub{\tau}{ij},\ve{B},\sigma^2,\nu)
p(\vesub{\tau}{ij}|\vesub{\theta}{ij})d\vesub{\tau}{ij}}}.\nonumber
\end{align}
However, the marginal log-likelihood can not be computed directly, because of intractable integration.
To avoid the complicated integration, some methods, such as Laplace approximation and the Markov Chain Monte Carlo(MCMC) method, are used to approximate  the marginal log-likelihood. But the error rate of the Laplace approximation may be $O(1)$ as the dimension of $\vesub{\tau}{ij}$ increases with the sample size $n$ \citep{r13}, and the MCMC method is too complicated to solve the problem conveniently, see Section 7.1.4 in \cite{r16}. This paper uses a Gaussian approximation method to compute $p(\ve{\mathcal{D}_n}|\ve{\theta},\ve{B},\nu)$ as follows
\begin{equation}\label{2.6}
p(\ve{\mathcal{D}_n}|\ve{\theta},\ve{B},\nu)\dot{=}
\prod\limits_{i,j}{\left.\frac{p(\vesub{Y}{ij},
\vesub{\tau}{ij}|\vesub{\theta}{ij},\ve{B},\sigma^2,\nu)}{p_G(\vesub{\tau}{ij}
|\vesub{Y}{ij},\vesub{\theta}{ij},\ve{B},\sigma^2,\nu)}\right|_{\vesub{\tau}{ij}=\hvesub{\tau}{ij}}},
\end{equation}
where $p_G(\vesub{\tau}{ij}|\vesub{Y}{ij},\vesub{\theta}{ij},\ve{B},\sigma^2,\nu)$ is the Gaussian approximation to the full conditional density $p(\vesub{\tau}{ij}|\vesub{Y}{ij},\vesub{\theta}{ij},\ve{B},\sigma^2,\nu)$ and $\hvesub{\tau}{ij}$ is the mode of the full conditional density for given $\vesub{\theta}{ij}$,~$\ve{B},~\sigma^2$ and $\nu$. Let $\tau_{ijk}=\tau_{ij}(t_k)$. We have
\[p(\vesub{Y}{ij},\vesub{\tau}{ij}|\vesub{\theta}{ij},\ve{B},\sigma^2,\nu)=
\text{exp}\{\text{log}(p(\vesub{\tau}{ij}|\vesub{\theta}{ij}))
+\sum\limits_{k=1}^n{\text{log}(p(y_{ij}(t_k)|\tau_{ijk},\ve{B},\sigma^2,\nu))}\}.\]
From Taylor expansion, we approximate $g_{ijk}(\tau_{ijk})\triangleq \text{log}(p(y_{ij}(t_k)|\tau_{ijk},\ve{B},\sigma^2,\nu))$ to the second order
\[g_{ijk}(\tau_{ijk})\approx g_{ijk}(\tau_{ijk}^{(0)})+\alpha_{ijk}\tau_{ijk}-\frac{1}{2}d_{ijk}\tau_{ijk}^2\]
where
\[\alpha_{ijk}=g_{ijk}^{'}(\tau_{ijk}^{(0)})-g_{ijk}^{''}(\tau_{ijk}^{(0)})\tau_{ijk}^{(0)},~~d_{ijk}=-g_{ijk}^{''}(\tau_{ijk}^{(0)}).\]
Let $S_{ijk}=y_{ij}(t_k)-(\vesup{\Phi}{\top}(t_i)\otimes\vess{X}{ij}{\top}) \text{Vec}(\vesup{B}{\top})-\tau_{ijk}.$
From density function of EMTD, it gives
\[g_{ijk}^{'}(\tau_{ijk})=\frac{(1+2\nu)S_{ijk}}
{2(\nu-1)\sigma^2+S_{ijk}^2},~~g_{ijk}^{''}(\tau_{ijk})
=\frac{(1+2\nu)(S_{ijk}^2-2(\nu-1)\sigma^2)}{(2(\nu-1)\sigma^2+S_{ijk}^2)^2}.\]
Therefore,
\[p(\vesub{Y}{ij},\vesub{\tau}{ij}|\vesub{\theta}{ij},\ve{B},\sigma^2,\nu) \propto \text{exp}\left\{-\frac{1}{2}\vess{\tau}{ij}{\top}\vess{K}{ij}{-1}\vesub{\tau}{ij}-
\frac{\vess{\tau}{ij}{\top}\vesub{D}{ij}\vesub{\tau}{ij}}{2}+\vess{\alpha}{ij}{\top}
\vesub{\tau}{ij}\right\},\]
where $\vesub{D}{ij}=\text{diag}(d_{ij1},\ldots,d_{ijn})$ and $\vess{\alpha}{ij}{\top}=(\alpha_{ij1},\ldots,\alpha_{ijn})$.

Next, for some initial value $\vess{\tau}{ij}{(0)}$, we can use Fisher scoring algorithm to find the Gaussian approximation, and the $k$th iteration is given by

(\romannumeral 1) Find the solution $\vess{\tau}{ij}{(k)}$ from $(\vess{K}{ij}{-1}+\vesub{D}{ij})\vess{\tau}{ij}{(k)}=\vesub{\alpha}{ij}$;

(\romannumeral 2) Update $\vesub{\alpha}{ij}$ and $\vesub{D}{ij}$ using $\vess{\tau}{ij}{(k)}$ and repeat (\romannumeral 1 ).\\
After the process converges, say at $\hvesub{\tau}{ij}$, we can get the Gaussian approximation $p_G(\vesub{\tau}{ij}|\vesub{Y}{ij},\vesub{\theta}{ij},\ve{B},\sigma^2,\nu)$ which is the density function of the normal distribution $MVN(\hvesub{\tau}{ij},(\vess{K}{ij}{-1}+\vesub{D}{ij})^{-1})$.

Considering smoothness of $\ve{\beta}(\cdot)$, we need to add a penalty $\text{PEN}_{\lambda}(\ve{\beta})$ into the log-likelihood function, denoted by
\begin{equation}\label{2.7}
l(\ve{B},\ve{\theta},\nu)=\sum\limits_{m=1}^M{\text{log}\int{p(\vesub{Y}{ij}
|\vesub{\tau}{ij},\ve{B},\sigma^2,\nu)
p(\vesub{\tau}{ij}|\vesub{\theta}{ij})d\vesub{\tau}{ij}}}+\text{PEN}_\lambda(\ve{\beta}),
\end{equation}
where $\lambda$ is a tuning parameter. This paper takes the penalized function,
\begin{equation}\label{2.8}
\begin{aligned}
\text{PEN}_\lambda(\ve{\beta})&=\lambda{\int_a^b{\|L\ve{\beta}(t)\|^2dt}}\\
&=\lambda\sum\limits_{i=1}^{I+1}{\text{trace}[((\vesup{B}{\top})_i)^\top(\vesup{B}{\top})_iL_{\Phi\Phi}]},
\end{aligned}
\end{equation}
where $L_{\Phi\Phi}=\int_a^b{[L\ve{\Phi}(t)][L\ve{\Phi}(t)^\top]dt}$ and $L$ is a derivative operator.
The form of $L$ depends on the property of function $\ve\beta(t)$, such as for the second order continuously derivative function,
$L\ve{\Phi}(t)=\omega_0\ve{\Phi}(t)+\omega_1D\ve{\Phi}(t)+D^2\ve{\Phi}(t)$,
where the weights $\omega_0,\omega_1$ may be either constant or function $\omega_0(t),\omega_1(t)$, and $D\ve{\Phi}(t),D^2\ve{\Phi}(t)$ mean the first and second derivative of $\ve{\Phi}(t)$.

From (\ref{2.6}), (\ref{2.7}) and (\ref{2.8}),we obtain an estimation procedure for parameters $\ve{\theta}$, $\ve{B}$ and $\nu$. For initial values $\ve{\theta}=\vesub{\theta}{0}$, $\ve{B}=\vesub{B}{0}$  and $\nu=\nu_{0}$, \\
(\uppercase\expandafter{\romannumeral 1}) For given $\ve{\theta}$, $\ve{B}$ and $\nu$, obtain$\hve{\tau}=(\hvesub{\tau}{1n},\ldots,\hvesub{\tau}{Mn})$ by the Fisher scoring algorithm above-mentioned.\\
(\uppercase\expandafter{\romannumeral 2}) For given $\ve{\tau}=\hve{\tau}$, we update estimates of $\ve{\theta}$, $\vesup{B}{\top}$ and $\nu$  by maximizing the penalized log-likelihood (\ref{2.7}) with (\ref{2.6}) and (\ref{2.8}).\\
(\uppercase\expandafter{\romannumeral 3}) Repeat (\uppercase\expandafter{\romannumeral 1})-(\uppercase\expandafter{\romannumeral 2}) until some convergence criteria hold.

\setcounter{equation}{0}

\section{Asymptotic properties}

\subsection{Robustness}

Influence function is an important index to measure the robustness of estimation method to outliers.
It is known that Gaussian distribution to fit data tends to be susceptive to outliers and has unbounded
influence function, while t-distribution has heavy-tail and bound influence function. This paper shows that the proposed functional ANOVA models with t-process error (TP model) is more robust than those with Gaussian process error (GP model), see the next theorem (its proof seen in Appendix A).

\vskip 10pt
\noindent{\bf Theorem 1.} {\it
If kernel function $k_{ij}$ is bounded for the $j$-th curve under level $i$, $j=1,...,J_i,~i=1,...,I$, and continuously differentiable on $\vesub{\theta}{ij}$, then for given $\nu$, the estimators $\hve{B}$ and $\hve{\theta}$ from the TP model have bound influence functions, while that from the GP model does not.
}

\subsection{Information consistency}

The consistency related to the estimation approach for model (\ref{2.2}) involves two aspects: the common mean parameter function $\ve{\beta}(t)$ and the random effect curve $\tau_{ij}(t)$ (or the curve $y_{ij}(t)$). The common mean function is estimated by collected data from all curves and its consistency is already proved in many different functional linear models under suitable conditions, see \cite{r20}, \cite{r11}. Therefore, this paper focuses on the second issue, the information consistency of $\hat{y}_{ij}(\cdot)$ to $y_{ij}(\cdot)$. Information consistency of the estimation method is also an important issue. For Gaussian process regression model and the extended t-process regression model, this properties are obtained in \cite{r14}, \cite{r17} and \cite{r18}. Here, we show this property for model (\ref{2.2}).

Without loss of generality, suppose the mean function about the $j-$th curve under level $i$, denoted by $\vesub{\mu}{n}\triangleq(\vess{X}{ij}{\top}\otimes\vess{\Phi}{n}{\top})\text{Vec}(\ve{B})$, has already been known. Let covariate values $\ve{U}_n=\{\vesub{u}{ij}(t_1),\ldots,\vesub{u}{ij}(t_n)\} $ where $\vesub{u}{ij}(t_k) \in \mathcal{X} \subset R^p$ are independently drawn from a distribution $\mathcal{U}(\ve{x})$. Let $p_\theta(\tau)$ be a measure induced by the process~$\tau(\cdot)$ on space $\mathcal{F}=\{\tau(\cdot):\mathcal{X}\rightarrow R\}.$

Denote
\[p_{TP}(\vesub{Y}{ij})=\int_{\mathcal{F}}{p(y_{ij}(t_1),...,y_{ij}(t_n)|{{\tau}_{ij}})dp_\theta({\tau}_{ij})},\]
\[p_{0}(\vesub{Y}{ij})=p(y_{ij}(t_1),...,y_{ij}(t_n)|{{\tau}_{0}}),\]
where $p_{TP}(\vesub{Y}{ij})$ is the Bayesian predictive distribution of $\vesub{Y}{ij}$ based on the t-process functional ANOVA model, and $p_{0}(\vesub{Y}{ij})$ is based on the true underlying function ${\tau}_{0}$ of $\tau_{ij}$. It is said that $p_{TP}(\vesub{Y}{ij})$ achieves information consistency if
\begin{equation}\label{3.1}
\frac{1}{n}E_{\vesub{U}{n}}(D[p_{0}(\vesub{Y}{ij}),p_{TP}(\vesub{Y}{ij})])\rightarrow 0~as~n\rightarrow \infty,
\end{equation}
where $E_{\vesub{U}{n}}$ denotes the expectation under the distribution of $\vesub{U}{n}$ and $D[f,g]$ is the Kullback-Leibler divergence between functions $f$ and $g$, that is,
\begin{equation}\label{3.2}
D[f,g]=\int{f(t)\text{log}\frac{f(t)}{g(t)}dt.}
\end{equation}
\vskip 10pt
\noindent{\bf Theorem 2.} {\it
Under the appropriate conditions in Lemma 1 and condition (A) in Appendix B, we have for $j=1,\ldots,J_i,~i=1,...,I$
\begin{equation}\label{3.3}
\frac{1}{n}E_{\vesub{U}{n}}(D[p_{0}(\vesub{Y}{ij}),p_{TP}(\vesub{Y}{ij})])\rightarrow 0~as~n\rightarrow \infty,
\end{equation}
where the expectation is taken over the distribution $\vesub{U}{n}$.
}

The proof of Theorem 2 is in Appendix B.

\section{Numerical studies}

\subsection{Simulation studies}

Numerical simulations are constructed to evaluate the robustness of the t-process functional ANOVA model (\ref{2.2}) (TP). The performance of TP is investigated by comparison of model (\ref{1.1}) with error term having Gaussian process (GP). In addition, we also compute model (\ref{2.2}) ignoring the random effect $\tau_{ij}$, denoted by TP($\tau=0$).

Simulation data are generated from three different models:
\begin{itemize}
\item
Model 1: The ANOVA model (\ref{1.1}) with $I=3$ and $J=2$, where $\mu(t)=t^2$, $\alpha_1(t)=0.5\cos(t)$,
$\alpha_2(t)=0.1\cos(t)$ and $\alpha_3(t)=-0.6\cos(t)$, $u_{ij}(t)=0.2t$ and $\theta=(0.1,10,0.1)$, $\varepsilon_{ij} \sim ETP(1.1,0.1,0,0.1)$.
\item
Model 2: $\varepsilon_{ij} \sim GP(0,0.1)$, the other setups are the same as Model 1.
\item
Model 3: $\tau_{ij}=0$, the other setups are the same as Model 1.
\end{itemize}
Model 1 is the true model for the TP method, while Model 2 is the true model for the GP method. Model 3 does not consider the random effect. Let observed time $t$ takes  61 points evenly spaced in [0,2]. Take $n$ sample points as training data with $n=11$ and $21$, the left are test data. Furthermore, to illustrate the performance of robustness, we randomly choose one point from the training data and disturb it by adding an extra error: a constant error 2 or random errors: $N(0,2)$ and $t(3)$, where $N(0,2)$ is normal distribution with mean 0 and variance 2, and $t(3)$ is student $t$ distribution with degree of freedom 3. All simulation are repeated 500 times.

Prediction errors (PE) and mean squared errors (MSE) of prediction from the TP, GP and TP($\tau=0$) methods  are presented in Tables 1 - 3 for Model 1 - 3, respectively. From these tables, it follows that the TP method has the smallest PE and MSE, while the GP has comparable results with
the TP($\tau=0$). It shows that the TP has more robustness against outliers than the GP, and better prediction than the TP($\tau=0$).

\begin{table}\label{tab1}
\centering
\caption{Prediction errors and mean squared errors of prediction results from the TP, GP and TP($\tau=0$) methods for generated data from Model 1, where their standard deviations are in parentheses.}
\vskip 18pt
\tabcolsep=15pt \fontsize{10}{18}\selectfont
\centerline{\tabcolsep=3truept\begin{tabular}{cccccc} \hline
\multicolumn{1}{c}{n}
         & \multicolumn{1}{c}{disturb error}
         & \multicolumn{1}{c}{}
         & \multicolumn{1}{c}{TP}
         & \multicolumn{1}{c}{GP}
         & \multicolumn{1}{c}{TP($\tau=0$)} \\
\hline
11&2&PE&0.5777(0.3670)&0.6615(0.3769)&0.6613(0.3768)\\
  &&MSE&0.5047(0.2484)&0.5886(0.2450)&0.5884(0.2450)\\
  &N(0,2)&PE&0.5516(0.2422)&0.6100(0.2524)&0.6098(0.2524)\\
  &&MSE&0.4785(0.1821)&0.5369(0.1910)&0.5367(0.1909)\\
  &t(3)&PE&0.4750(0.4126)&0.5173(0.4166)&0.5172(0.4167)\\
  &&MSE&0.4137(0.3972)&0.4560(0.4018)&0.4559(0.4018)\\
\hline
21&2&PE&0.3952(0.1784)&0.4267(0.1768)&0.4258(0.1767)\\
    &&MSE&0.3288(0.1311)&0.3604(0.1239)&0.3595(0.1236)\\
   &N(0,2)&PE&0.3676(0.1362)&0.3976(0.1358)&0.3970(0.1358)\\
         &&MSE&0.3077(0.1146)&0.3376(0.1147)&0.3370(0.1147)\\
   &t(3)&PE&0.3630(0.2764)&0.3888(0.2689)&0.3883(0.2687)\\
       &&MSE&0.3004(0.2591)&0.3261(0.2513)&0.3256(0.2511)\\
\hline
\end{tabular}}
\end{table}

\begin{table}\label{tab2}
\centering
\caption{Prediction errors and mean squared errors of prediction results from the TP, GP and TP($\tau=0$) methods for generated data from Model 2, where their standard deviations are in parentheses.}
\vskip 18pt
\tabcolsep=15pt \fontsize{10}{18}\selectfont
\centerline{\tabcolsep=3truept\begin{tabular}{cccccc} \hline
\multicolumn{1}{c}{n}
         & \multicolumn{1}{c}{disturb error}
         & \multicolumn{1}{c}{}
         & \multicolumn{1}{c}{TP}
         & \multicolumn{1}{c}{GP}
         & \multicolumn{1}{c}{TP($\tau=0$)} \\
\hline
11&2&PE&0.6209(0.1621)&0.7231(0.1598)&0.7229(0.1596)\\
  &&MSE&0.5196(0.1600)&0.6217(0.1574)&0.6215(0.1573)\\
  &N(0,2)&PE&0.6097(0.1976)&0.6892(0.2089)&0.6890(0.2089)\\
  &&MSE&0.5107(0.1955)&0.5900(0.2081)&0.5898(0.2080)\\
  &t(3)&PE&0.5549(0.3328)&0.6163(0.3331)&0.6162(0.3331)\\
  &&MSE&0.4563(0.3329)&0.5179(0.3330)&0.5178(0.3330)\\
\hline
21&2&PE&0.4516(0.1275)&0.4864(0.1246)&0.4857(0.1243)\\
    &&MSE&0.3514(0.1238)&0.3863(0.1203)&0.3855(0.1200)\\
   &N(0,2)&PE&0.4237(0.1340)&0.4572(0.1306)&0.4568(0.1303)\\
         &&MSE&0.3235(0.1316)&0.3570(0.1279)&0.3565(0.1276)\\
   &t(3)&PE&0.4104(0.1775)&0.4492(0.1712)&0.4487(0.1711)\\
       &&MSE&0.3103(0.1782)&0.3492(0.1713)&0.3488(0.1712)\\
\hline
\end{tabular}}
\end{table}

\begin{table}\label{tab3}
\centering
\caption{Prediction errors and mean squared errors of prediction results from the TP, GP and TP($\tau=0$) methods for generated data from Model 3, where their standard deviations are in parentheses.}
\vskip 18pt
\tabcolsep=15pt \fontsize{10}{18}\selectfont
\centerline{\tabcolsep=3truept\begin{tabular}{cccccc}
\hline
\multicolumn{1}{c}{n}
         & \multicolumn{1}{c}{disturb error}
         & \multicolumn{1}{c}{}
         & \multicolumn{1}{c}{TP}
         & \multicolumn{1}{c}{GP}
         & \multicolumn{1}{c}{TP($\tau=0$)} \\
\hline
11&2&PE&0.3951(0.1547)&0.4758(0.1429)&0.4755(0.1429)\\
  &&MSE&0.3363(0.1313)&0.4171(0.1161)&0.4168(0.1161)\\
  &N(0,2)&PE&0.4023(0.2059)&0.4492(0.2124)&0.4490(0.2124)\\
  &&MSE&0.3426(0.1881)&0.3895(0.1950)&0.3893(0.1950)\\
  &t(3)&PE&0.3876(0.2356)&0.4203(0.2352)&0.4201(0.2351)\\
  &&MSE&0.3243(0.2262)&0.3571(0.2260)&0.3569(0.2259)\\
  \hline
  21&2&PE&0.2260(0.1253)&0.2621(0.1172)&0.2608(0.1171)\\
    &&MSE&0.1650(0.0766)&0.2011(0.0653)&0.1999(0.0651)\\
   &N(0,2)&PE&0.2098(0.1331)&0.2439(0.1310)&0.2432(0.1309)\\
         &&MSE&0.1461(0.0841)&0.1802(0.0804)&0.1795(0.0802)\\
   &t(3)&PE&0.1989(0.2559)&0.2225(0.2526)&0.2220(0.2525)\\
       &&MSE&0.1338(0.2068)&0.1575(0.2029)&0.1570(0.2028)\\
\hline
\end{tabular}}
\end{table}

\subsection{Real data examples}

\begin{figure}[htbp]
  \centering
  \includegraphics[width=0.9\textwidth]{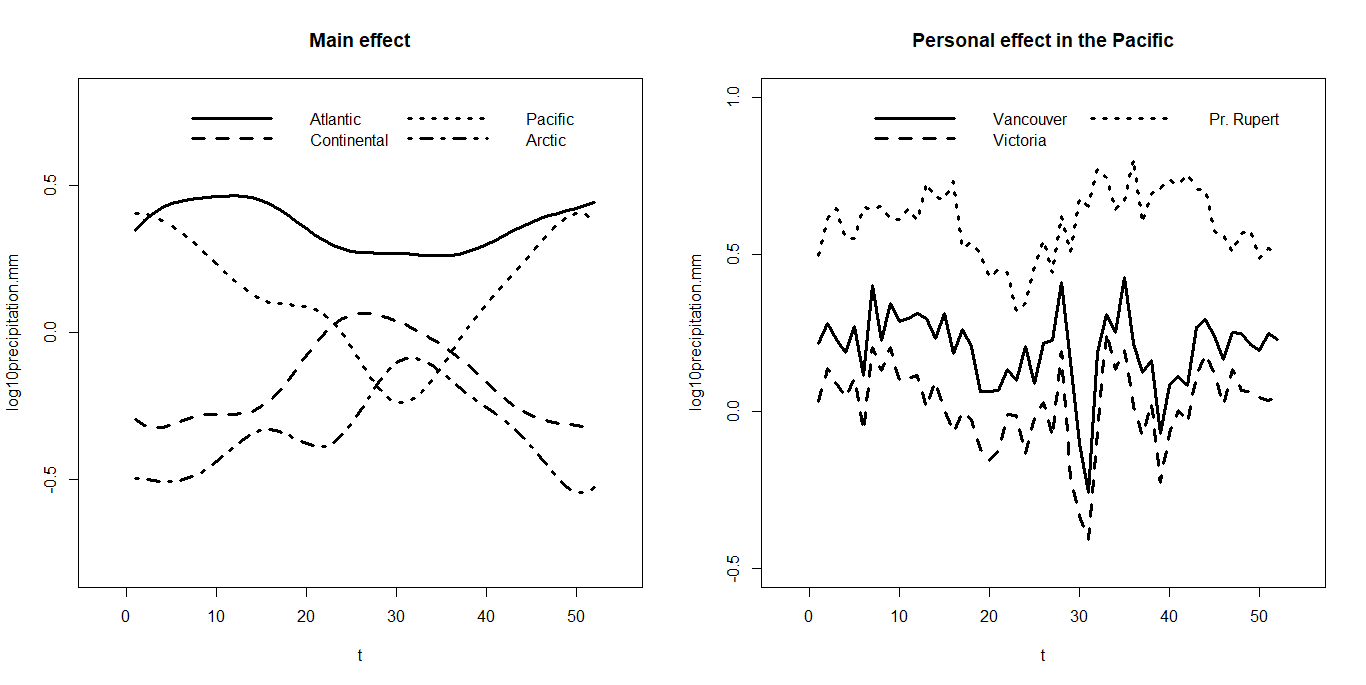}
  \caption{The Main effect and examples of the personal effect of the Canada weather data.}\label{fig:digit}
\end{figure}
\begin{figure}[htbp]
  \centering
  \includegraphics[width=0.9\textwidth]{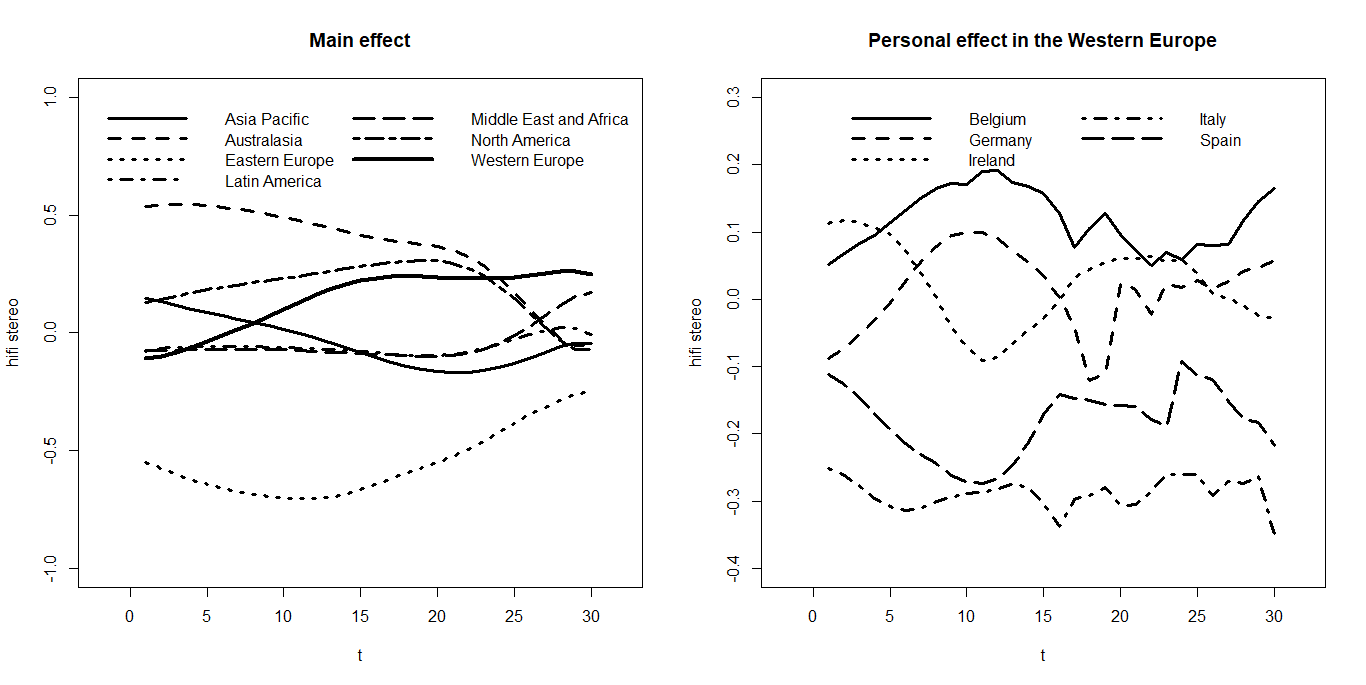}
  \caption{The Main effect and examples of the personal effect of the Hi-Fi stereo data.}\label{fig:digit}
\end{figure}
\begin{figure}[htbp]
  \centering
  \includegraphics[width=0.9\textwidth]{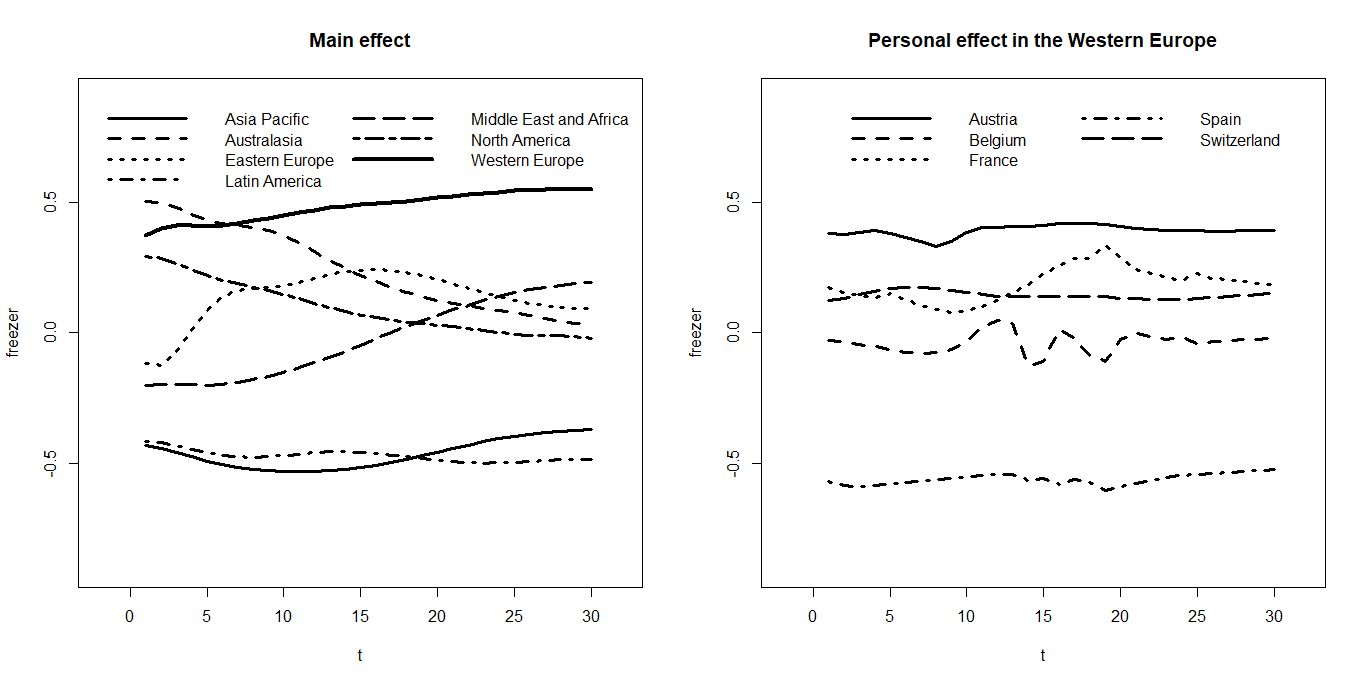}
  \caption{The Main effect and examples of the personal effect of the freezer data.}\label{fig:digit}
\end{figure}

The robust functional ANOVA model is applied to two datasets, Canada weather data and  market penetration of new products. Canadian weather data includes temperature and precipitation of weather  at 35 Canadian weather observation stations, which is available in the R-package fda \citep{r10}. These stations are divided into 4 regions: Arctic, Atlantic, Continental and Pacific. We study the week-average precipitation for 4 regions, where each region is a level  and  the week-average temperature is variable for random effect $\tau$. For the market penetration data \citep{r1}, we take the market penetration of Hi-Fi stereo and freezer as examples to illustrate our method. There are 86 countries in 7 continents from 1977 to 2015. Due to the missing data, we finally take data of 36 countries from 1983 to 2012 for the Hi-Fi stereo, and 35 countries from 1986 to 2015 for the freezer. With respect to relation with the Hi-Fi stereo and freezer, for random effect $\tau$, we take the penetration of personal computer as variable for the Hi-Fi stereo, and that of microwave oven for the freezer. Because 36 countries come from seven regions: Asia Pacific, Australasia, Eastern Europe, Latin America, Middle East and Africa, North America and Western Europe. So, the number of levels is 7.

Figures 1, 2 and 3 present the main effects and examples of the personal effect of some subjects for the weather data, the market penetration of Hi-Fi stereo and freezer, respectively. We can see that besides different main effects for the studied levels,  each subject (weather station or country) also has different personal character.

For these two datasets, we respectively randomly choose 30 and 20 time points as training data and the rest as test data. Three estimation methods, TP, GP and TP($\tau=0$), are applied to fit the training data and then predict the test data. These procedures are repeated for 500 times. Table 4 shows prediction errors from TP, GP and TP($\tau=0$). We can see the TP has the smallest prediction errors,
while GP and TP($\tau=0$) have the comparable ones.

\begin{table}\label{tab4}
  \centering
  \caption{Prediction errors of Canada precipitation data, market penetrations of the freezer and Hi-Fi stereo from TP, GP and TP($\tau=0$),
  where standard deviation is in parentheses.}
  \vskip 18pt
  \tabcolsep=15pt \fontsize{10}{18}\selectfont
  \begin{tabular}{l cccc  ccc}
     \hline\noalign{\smallskip}
      {Data} &{TP} &{GP}&{TP($\tau=0$)}\\
     \hline\noalign{\smallskip}
     Precipitation &0.0416(0.0029)&0.0488(0.0046)&0.0490(0.0045)\\
     Freezer &0.1362(0.0060)&0.1480(0.0057)&0.1487(0.0055)\\
     Hi-Fi stereo &0.0583(0.0149)&0.0641(0.0144)&0.0645(0.0144)\\
     \hline\noalign{\smallskip}
   \end{tabular}
\end{table}

\section{Conclusion}

The functional ANOVA model with Gaussian process offers a model for data with multi-dimensional covariances, and the specification of covariance kernels gives the ability to analyze a wide class of different type data to the model. However, the model with Gaussian process is not robust to outliers. Some heavy-tail processes can be used to overcome this shortcoming. This paper uses Gaussian process and t process to fit the personal random effect and error term, respectively, which results in a robust functional ANOVA model with t-process. Robustness property and information consistency of the proposed method are obtained. Numerical examples show that the robust functional ANOVA model has more robustness than that with Gaussian process.

Furthermore, it is a natural way to use t process to fit the random effect instead of Gaussian process, which leads to both of the personal random effect and error term having t processes. This extension makes estimation procedure more complicated because of more intractable integration.

\appendix

\setcounter{equation}{0}

\section{Robustness}

\noindent{\bf{ Proof of Theorem 1:}}\\
From the log-likelihood function $l(\ve{\theta},\ve{B},\nu)$ defined in (\ref{2.7}), we get the score functions of $\text{Vec}(\vesup{B}{\top})$,$~\vesub{\theta}{ij}$,$~\sigma^2$ as follows,
\begin{align}
\leftline{$s(\text{Vec}(\vesup{B}{\top});\ve{\mathcal{D}_n})$}\nonumber\\
\leftline{$=\sum\limits_{i,j}{\left(\sum\limits_{k=1}^n{\frac{(1+2\nu)S_{ijk}(\ve{\Phi}(t_i)\otimes\vesub{X}{ij}) }{2(\nu-1)\sigma^2+S_{ijk}^2}}-\frac{1}{2}\text{tr}((\vess{K}{ij}{-1}+\vesub{D}{ij})^{-1}\ve{A})\right)}+2\lambda \text{Vec}(\vesup{B}{\top}L_{\Phi\Phi}),$}\nonumber
\end{align}
\begin{align}
\leftline{$s(\vesub{\theta}{ij};\ve{\mathcal{D}_n})$}\nonumber\\
\leftline{$=\frac{1}{2}\text{tr}((\vess{K}{ij}{-1}(\vess{K}{ij}{-1}+\vesub{D}{ij})^{-1}
\vess{K}{ij}{-1}+\vess{K}{ij}{-1}\hvesub{\tau}{ij}\hvess{\tau}{ij}{\top}\vess{K}{ij}{-1}
-\vess{K}{ij}{-1})\frac{\partial\vesub{K}{ij}}{\partial\theta_{ijk}})_{1\times(2p+1)},
$}\nonumber\\
\leftline{$s(\sigma^2;\ve{\mathcal{D}_n})$}\nonumber\\
\leftline{$=\sum\limits_{i,j}{\left(
-\frac{n}{2\sigma^2}+\sum\limits_{k=1}^n{(\frac{(1+2\nu)(\nu-1)S_{ijk}^2}{(2(\nu-1)\sigma^2)^2+2(\nu-1)\sigma^2S_{ijk}^2})}-
\frac{1}{2}\text{tr}((\vess{K}{ij}{-1}+\vesub{D}{ij})^{-1}\frac{\partial\vesub{D}{ij}}{\partial\sigma^2})\right)},$}\nonumber
\end{align}
where for $q=1,\ldots,p, p=1,\ldots,L\times(I+1),k=1,\ldots,n$,
\begin{align}
&\ve{A}=\text{diag}\left(\frac{2(1+2\nu)S_{ijk}(6(\nu-1)\sigma^2-S_{ijk}^2)
(\vesup{\Phi}{\top}(t_i)\otimes\vess{z}{ij}{\top})_{p}}
{(2(\nu-1)\sigma^2+S_{ijk}^2)^3}\right),\nonumber\\
&\frac{\partial\vesub{D}{ij}}{\partial\sigma^2}=\text{diag}(\frac{2(1+2\nu)(\nu-1)(3S_{ijk}^2-2(\nu-1)\sigma^2)}
{(2(\nu-1)\sigma^2+S_{ijk}^2)^3}),\nonumber
\end{align}
\begin{align}
\frac{\partial\vesub{K}{ij}}{\partial\theta_{ij0}}&=\left(\text{exp}\left\{
-\sum\limits_{k=1}^p{\frac{\theta_{ijk}(u_{ijk}(t_l)-u_{ijk}(t_m))^2}{2}}\right\}\right)_{n\times n},\nonumber\\
\frac{\partial\vesub{K}{ij}}{\partial\theta_{ijq}}&=\left(\theta_{ij0}\text{exp}\left\{
-\sum\limits_{k=1}^p{\frac{\theta_{ijk}(u_{ijk}(t_l)-u_{ijk}(t_m))^2}{2}}\right\}(
-\frac{(u_{ijq}(t_l)-u_{ijq}(t_m))^2}{2})\right)_{n\times n}
,\nonumber\\
\frac{\partial\vesub{K}{ij}}{\partial\eta_{ijq}}&=\left(u_{ijq}(t_l)u_{inq}(t_m)\right)_{n\times n}
(q=1,\ldots,p).\nonumber
\end{align}
So, we can get that when $\vesub{Y}{ij} \rightarrow \infty$ for $j$-th object under level $i$, the score functions of $\text{Vec}(\vesup{B}{\top})$, $~\vesub{\theta}{ij}$ and $~\sigma^2$ are bounded, while that from the GPR model tend to $\infty$.

Let $T(F_n)=T_n(\vesub{Y}{11},\ldots,\vesub{Y}{IJ_I})$ be an estimate of $\ve{\beta}$ (in this subsection let $\ve{\beta}=\ve{\theta},\ve{B}$), in which $F_n$ is the empirical distribution of $\{\vesub{Y}{11},\ldots,\vesub{Y}{IJ_I}\}$ and $T$ is a functional on some subset of all distributions. Let $\ve{Y}=(\vess{Y}{11}{\top},\ldots,$ $\vess{Y}{IJ_I}{\top})$, according to \cite{r5}, influence function of $T$ at $F$ is defined as
\[IF(\ve{y};T,F)=\lim_{t \rightarrow 0}\frac{T((1-t)F+t\delta_{\ve{y}})-T(F)}{t},\]
where $\delta_{\ve{y}}$ put mass 1 on point $\ve{y}$ and 0 on others.

Following \cite{r5} , we can get that estimator $\hve{\beta}$ of $\ve{\beta}$ has the influence function
\[
IF(\ve{Y};\hve{\beta},F)=-(E(\frac{\partial ^2l(\ve{\theta},\ve{B},\nu)}{\partial \ve{\beta}
\partial \vesup{\beta}{\top}}))^{-1}s(\ve{\beta};\ve{Y}).
\]

The matrix $\frac{\partial ^2l(\ve{\theta},\ve{B},\nu)}{\partial \ve{\beta}
\partial \vesup{\beta}{\top}}$ is bounded according to $\vesub{Y}{ij},i=1,\ldots,I,~j=1,\ldots,J_i$, therefore the influence functions of $\hve{\beta}$ are bounded under the propsoed functional ANOVA model. However, those under the GPR model are unbounded. $\sharp$

\setcounter{equation}{0}

\section{Information consistency}

\newtheorem{lemma}{\textbf{Lemma}}
\begin{lemma}
Suppose $\vesub{Y}{ij}$ are independent samples from model (\ref{2.2}) and $\tau_0 \in \mathcal{F}$ has a Gaussian process with zero mean and bounded kernel function $k(\cdot,\cdot;\vesub{\theta}{ij})$. Let $k(\cdot,\cdot;\vesub{\theta}{ij})$ be continuous in $\vesub{\theta}{ij}$ and $\hvesub{\theta}{ijn}$ be one consistent estimator of $ \vesub{\theta}{ij}$. Then we have
\begin{align}
&-\text{log}p_{TP}(y_{ij}(t_1),\ldots,y_{ij}(t_n))+\text{log}p_0(y_{ij}(t_1),\ldots,y_{ij}(t_n))\nonumber\\
\le& c+\frac{1}{2}\text{log}|\ve{I_n}+\sigma^{-2}\vesub{K}{ij}|+\frac{1}{2}\|\tau_0\|_k^2+\delta,\nonumber
\end{align}
where $\|\tau_0\|_k$ is the reproducing kernel Hilbert space(RKHS) norm of $\tau_0$ associated with $k(\cdot , \cdot ;\vesub{\theta}{ij})$, $\vesub{K}{ij}$ is covariance matrix of $\tau_0$ over $\vesub{U}{n}$, $\ve{I_n}$ is the $n \times n$ identity matrix and $c,\delta$ are some positive constants.
\end{lemma}
\vskip 10pt
\noindent{\bf{ Proof:}}~~Let $\mathcal{H}$ is the reproducing kernel Hilbert space(RKHS) associated with $k(\cdot , \cdot ;\vesub{\theta}{ij})$, and $\mathcal{H}_n$ is the span of $\{k(\cdot,\vesub{u}{ij}(t_k);\vesub{\theta}{ij}\}$, that is, $\mathcal{H}_n=\{f(\cdot):f(\ve{x})=\sum\nolimits_{k=1}^n{\alpha_k k(\ve{x},\vesub{u}{ij}(t_k);\vesub{\theta}{ij})},\text{for any} ~\alpha_k \in R \}$ . First, we assume that $\tau_0 \in \mathcal{H}_n$, then we can express $\tau_0(\cdot)$ as
\[
\tau_0(\cdot)=\sum\limits_{k=1}^n{\alpha_k k(\cdot,\vesub{u}{ij}(t_k);\vesub{\theta}{ij})}
\triangleq K(\cdot)\ve{\alpha},
\]
where $K(\cdot)=(k(\cdot,\vesub{u}{ij}(t_1);\vesub{\theta}{ij}),\ldots,k(\cdot,\vesub{u}{ij}(t_n);\vesub{\theta}{ij}))$ and $\ve{\alpha}=(\alpha_1,\ldots,\alpha_n)^{\top}$. With the property of RKHS, we can get $\|\tau_0\|_k^2=\vesup{\alpha}{\top}\vesub{K}{ij}\ve{\alpha}$, and $(\tau_0(\vesub{u}{ij}(t_1)),$ $\ldots,\tau_0(\vesub{u}{ij}(t_n)))^{\top}=\vesub{K}{ij}\ve{\alpha}$.
\par By the Fenchel-Legendre duality relationship, we can get
\begin{equation}\label{B.1}
E_{\overline{p}}[\text{log}p(y_{ij}(t_1),\ldots,y_{ij}(t_n))|\vesub{\tau}{ij}] \le \text{log} E_p[p(y_{ij}(t_1),\ldots,y_{ij}(t_n))|\vesub{\tau}{ij}]+D[\overline{p},p],
\end{equation}
where $p$ is the measure induced by $GP(0,k(\cdot,\cdot;\hvesub{\theta}{ij}))$, therefore its finite dimensional distribution at $\{y_{ij}(t_1),\ldots,y_{ij}(t_n)\}$ is $N(0,\hvesub{K}{ij})$, where $\hvesub{K}{ij}$ is defined in the same way as $\vesub{K}{ij}$ but with $\vesub{\theta}{ij}$ being replaced by its estimator $\hvesub{\theta}{ijn}$, and $\overline{p}$ is the posterior distribution of ${\tau}_{ij}$ with a prior distribution $GP(0,k(\cdot,\cdot;\vesub{\theta}{ij}))$ and normal likelihood $\prod\nolimits_{k=1}^n {N(\hat{y}_{ij}(t_k);{\tau}_{ij},\sigma^2)}$, where $\hvesub{y}{ij} \triangleq (\hat{y}_{ij}(t_1),\ldots,\hat{y}_{ij}(t_n))^{
\top}=(\vesub{K}{ij}+\sigma^2 \ve{I_n})\ve{\alpha}$.

Therefore, by the Gaussian process regression, the posterior distribution of ${\tau}_{ij}$ is
$\overline{p}({\tau}_{ij})=N(\vesub{K}{ij}\ve{\alpha},\vesub{K}{ij}\vesup{B}{-1})$, where $\ve{B}=\ve{I_n}+\sigma^{-2}\vesub{K}{ij}$. Then we have
\begin{equation}\label{B.2}
\begin{aligned}
D[\overline{p},p]&=\frac{1}{2}[-\text{log}|\hvess{K}{ij}{-1}\vesub{K}{ij}|+\text{log}|\ve{B}|+\text{tr}(\hvess{K}{ij}{-1}
\vesub{K}{ij}\vesup{B}{-1})\\
&+\|\tau_0\|_k^2+\vesup{\alpha}{\top}\vesub{K}{ij}(\hvess{K}{ij}{-1}\vesub{K}{ij}-\ve{I_n})
\ve{\alpha}-n].
\end{aligned}
\end{equation}

In addition, we show that
\begin{equation}\label{B.3}
 E_p[p(y_{ij}(t_1),\ldots,y_{ij}(t_n))|{\tau}_{ij}]=\int\nolimits_{\mathcal{F}}{p(y_{ij}(t_1),\ldots,y_{ij}(t_n)|{\tau}_{ij})
dp_{\theta}({\tau}_{ij})=p_{TP}(\vesub{Y}{ij})},
\end{equation}
and
\begin{align}\label{B.4}
&E_{\overline{p}}[\text{log}p(y_{ij}(t_1),\ldots,y_{ij}(t_n))|{\tau}_{ij}] \nonumber\\
&\ge \text{log}p_0(y_{ij}(t_1),\ldots,y_{ij}(t_n))-\frac{1}{2\sigma^2}\text{tr}(\vesub{K}{ij}\vesup{B}{-1}),\nonumber\\
&=\text{log}p_{0}(\vesub{Y}{ij})-\frac{1}{2\sigma^2}\text{tr}(\vesub{K}{ij}\vesup{B}{-1}).
\end{align}
Hence, it follow from (\ref{B.1}),~(\ref{B.2}),~(\ref{B.3})~and~(\ref{B.4}) that
\begin{align}\label{B.5}
&-\text{log}p_{TP}(y_{ij}(t_1),\ldots,y_{ij}(t_n))+\text{log}p_0(y_{ij}(t_1),\ldots,y_{ij}(t_n)) \nonumber\\
&\le\frac{1}{2}[-\text{log}|\hvess{K}{ij}{-1}\vesub{K}{ij}|+\text{log}|\ve{B}|+\text{tr}((\hvess{K}{ij}{-1}
\vesub{K}{ij}+\frac{1}{\sigma^2}\vesub{K}{ij})\vesup{B}{-1})+\|\tau_0\|_k^2\nonumber\\
&+\vesup{\alpha}{\top}\vesub{K}{ij}(\hvess{K}{ij}{-1}\vesub{K}{ij}-\ve{I_n})
\ve{\alpha}-n].
\end{align}

Since the covariance function is bounded and the covariance function is continous in $\vesub{\theta}{ij}$ and
$\hvesub{\theta}{ijn} \rightarrow \vesub{\theta}{ij}$, we have $\hvess{K}{ij}{-1}\vesub{K}{ij}-\ve{I_n}\rightarrow 0$ as $n\rightarrow \infty $. Hence, there exist
positive constants $c$ and $\varepsilon$ such that for $n$ large enough, $-\text{log}|\hvess{K}{ij}{-1}\vesub{K}{ij}|<c,$
$\vesup{\alpha}{\top}\vesub{K}{ij}(\hvess{K}{ij}{-1}\vesub{K}{ij}-\ve{I_n})
\ve{\alpha}<c,$ and $\text{tr}(\hvess{K}{ij}{-1}
\vesub{K}{ij}\vesup{B}{-1})<\text{tr}((\ve{I_n}+\varepsilon\vesub{K}{ij})\vesup{B}{-1}),$
which plugs in (\ref{B.5}), there exist positive constant $\delta$, we have the inequality
\begin{align}
\leftline{$-\text{log}p_{TP}(y_{ij}(t_1),\ldots,y_{ij}(t_n))+\text{log}p_0(y_{ij}(t_1),\ldots,y_{ij}(t_n))$}\nonumber\\
\leftline{$\le\frac{1}{2}[2c+\text{log}|\ve{B}|+\text{tr}((\ve{I_n}+(\varepsilon+\sigma^{-2})\vesub{K}{ij})\vesup{B}{-1}
)+\|\tau_0\|_k^2-n].$}\nonumber\\
\leftline{$\le c+\frac{1}{2}\text{log}|\ve{B}|+\frac{1}{2}\|\tau_0\|_k^2+\delta.$}\nonumber
\end{align}
From the Representer Theorem, it can obtain
\begin{align}
&-\text{log}p_{TP}(y_{ij}(t_1),\ldots,y_{ij}(t_n))+\text{log}p_0(y_{ij}(t_1),\ldots,y_{ij}(t_n))\nonumber\\
\le& c+\frac{1}{2}\text{log}|\ve{B}|+\frac{1}{2}\|\tau_0\|_k^2+\delta, \nonumber
\end{align}
for all $\tau_0(\cdot) \in \mathcal{H}$. The proof is over. $\sharp$

To prove Theorem 2, it need condition
\noindent (A) $\|\tau_0\|_k$ is bounded and $E_{\vesub{U}{n}}(\text{log}|\ve{I_n}+\sigma^{-2}\vesub{K}{ij}|)=o(n)$.

\ve{ Proof of Theorem 2.}\\
From the definition of the information consistency, we get that
\begin{align}
&D[p_{0}(\vesub{Y}{ij}),p_{TP}(\vesub{Y}{ij})]\nonumber\\
=&\int\nolimits_{\mathcal{F}}
{p_{0}(y_{ij}(t_1),\ldots,y_{ij}(t_n))\text{log}\frac{p_{0}(y_{ij}(t_1),\ldots,y_{ij}(t_n))}{p_{TP}(y_{ij}(t_1),\ldots,y_{ij}(t_n))}dy_{ij}(t_1)\cdots dy_{ij}(t_n)}\nonumber
\end{align}
From Lemma 1, we obtain that
\[
\frac{1}{n}E_{\vesub{U}{n}}(D[p_{0}(\vesub{Y}{ij}),p_{TP}(\vesub{Y}{ij})])\le
\frac{c}{n}+\frac{1}{2n}E_{\vesub{U}{n}}(\text{log}|\ve{I_n}+\sigma^{-2}\vesub{K}{ij}|)+\frac{1}{2n}\|\tau_0\|_k^2+\frac{\delta}{n},
\]
where $c$ and $\delta$ are two positive constants. Because $\|\tau_0\|_k$ is bounded and expected regret term~$E_{\vesub{U}{n}}(\text{log}|\ve{I_n}+\sigma^{-2}\vesub{K}{ij}|)=o(n)$, Theorem 2 holds. $\sharp$

\bibliographystyle{elsarticle-harv}

\end{document}